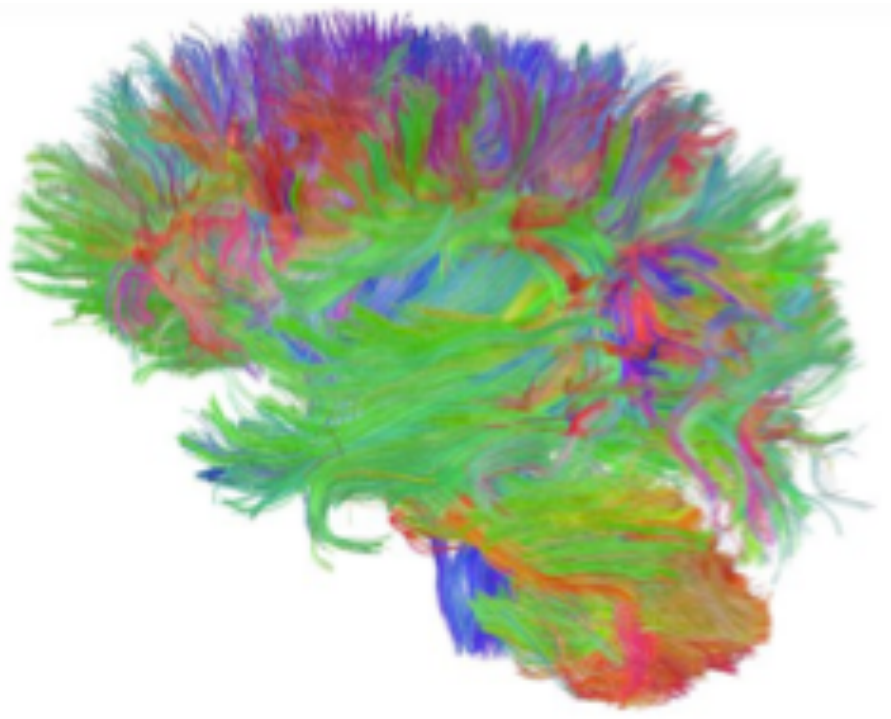

Marcus Kaiser

# Common Connectome Constraints: From *C. elegans* and *Drosophila* to *Homo sapiens*

Technical Report No. 4

Wednesday, 14 May 2014

Dynamic Connectome Lab

http://www.biological-networks.org/

# Common Connectome Constraints:
# From *C. elegans* and *Drosophila* to *Homo sapiens*

Marcus Kaiser[1, 2, 3]

[1] School of Computing Science, Newcastle University, UK

[2] Institute of Neuroscience, Newcastle University, UK

[3] Department of Brain and Cognitive Sciences, Seoul National University, South Korea

Corresponding author:

Dr Marcus Kaiser

School of Computing Science

Newcastle University

Claremont Tower, Newcastle upon Tyne, NE1 7RU, United Kingdom,

E-Mail: m.kaiser@ncl.ac.uk

## Introduction

Neuroscience, in a way, has always been a network science [1]. Studies over the last 20 years have used methods from graph theory [2] and network analyses [3] to understand the link between network structure and function (see Suppl. Note S1 for a glossary of terms). The nodes of a network can be neurons, populations of neurons, or brain regions, depending on the scale under examination. Connections can be chemical or electrical synapses, or fibre tracts and can be directed so that activity only travels in one direction (A → B) or bi-directional in that activity flows in both ways (A ↔ B). Directed edges play an important role in neural circuits as they allow for feed-forward and feed-back loops (Figure 1A). At the local level, directed edges can be formed through chemical synapses. At the global level, they can be formed through fibre tracts that project from one brain region to another but not vice versa. Most brain regions are connected in both directions, possibly providing direct feed-back or a top-down influence, but up to 15% of fibres are uni-directional [4]. These networks or graphs can be represented in an adjacency matrix where '1' represents an existing connection and '0' stand for a connection that either has not been discovered yet or that is known to be absent (Figure 1B). Alternatively, such matrices can indicate the weight of an existing connection using an ordinal scale (e.g., 1: weak; 2: medium; 3: strong) or metric scale (normalized value, say from 0 to 1). A matrix can also be plotted as a figure where the colour or greyscale indicates the strength of a connection (Figure 1C). Such figures are quite informative in distinguishing different kinds of networks, as we will see later on.

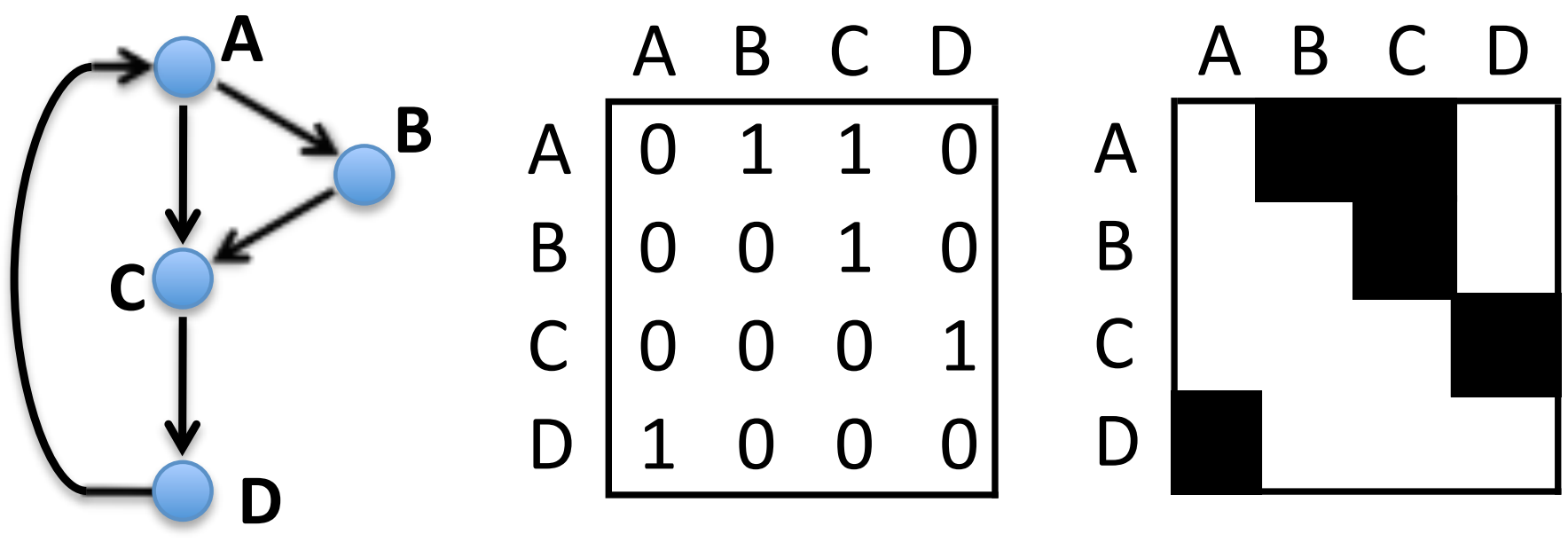

**Fig. 1.** Representing neural systems as networks. (left) Network with four nodes and feed-forward (A → C, A → B → C) and feed-back (A → C → D → A) loops. (middle) Representation in an adjacency matrix where existing connections result in a value of 1 and non-existent ones in a value of 0. (right) For visualisation, such binary matrices can be represented with black squares for existing and white squares for non-existing connections.

For neural networks, having an anatomical link is not the only type of connectivity [5]. Anatomical or physical connections between nodes form one type called *structural connectivity*. Alternatively, nodes can be linked if they show similar activity patterns forming *functional connectivity*. Such similarity could, for example, be measured in the correlation in the activity patterns of two brain regions or two neurons. The correlation value could either be used as strength of a connection or the correlation network could be binarized in that connection-weights are set to one if the corresponding correlation is above a certain threshold and zero otherwise. Finally, activity in one node can modulate (increase or decrease) activity in another node. In this case, there is a directed link going from the influencing node to the modulated node; networks of these links form the *effective connectivity*. A functional connection might indicate that two nodes are structurally connected but it might also arise if both nodes are driven by common input. However, a correlation between both types exist: in *C. elegans*, gene expression patterns correlate with the existence of axons between neurons [6] and in humans, structural and functional connectivity are linked [7].

**The evolution of neural networks**

As for other aspects of biology, it is useful to look at brain networks in terms of their evolution [8]. The first species to show neural networks are Coelenterates such as *Cnidaria*. These animals show a diffuse two-dimensional nerve net, which, in terms of network science, is called a regular or lattice network (Figure 2A). In such networks, neighbours are well connected but there are no long-distance connections. Such lattice networks are an important part of neural systems and still remain in two-dimensional lattices such as the retina as well as cortical and subcortical layered structures. For functionally specialised circuits, however, a regular organisation is unsuitable. Starting with the formation of sensory organs and motor units, neurons aggregate in ganglia. Such ganglia are often not only spatially clustered but also topologically clustered: topological clusters or modules (in social sciences communities) are sets of nodes with many connections within a module but few connections between modules (Figure 2B). In this way, ganglia can process one modality without interference from neurons processing different kinds of information. A well-studied example of a modular network of 302 neurons and 10 ganglia is the roundworm *Caenorhabditis elegans* [9,10], the first organism where the complete set of neural connections or 'connectome' [11,12] is known. At one point, having one module for one modality or function is not sufficient; an example is processing visual information in the rhesus monkey (macaque) where the visual module consists of two network components: nodes that form the dorsal pathway for processing object movement and nodes of the ventral pathway for processing objects features such as colour and form [13]. These networks where smaller sub-modules are nested within modules are called hierarchical networks (Figure 2C). Here, the largest modules would form the top level of the hierarchy whereas sub-modules within modules and sub-sub-modules within sub-modules are on the second and third level of the hierarchy, respectively [14].

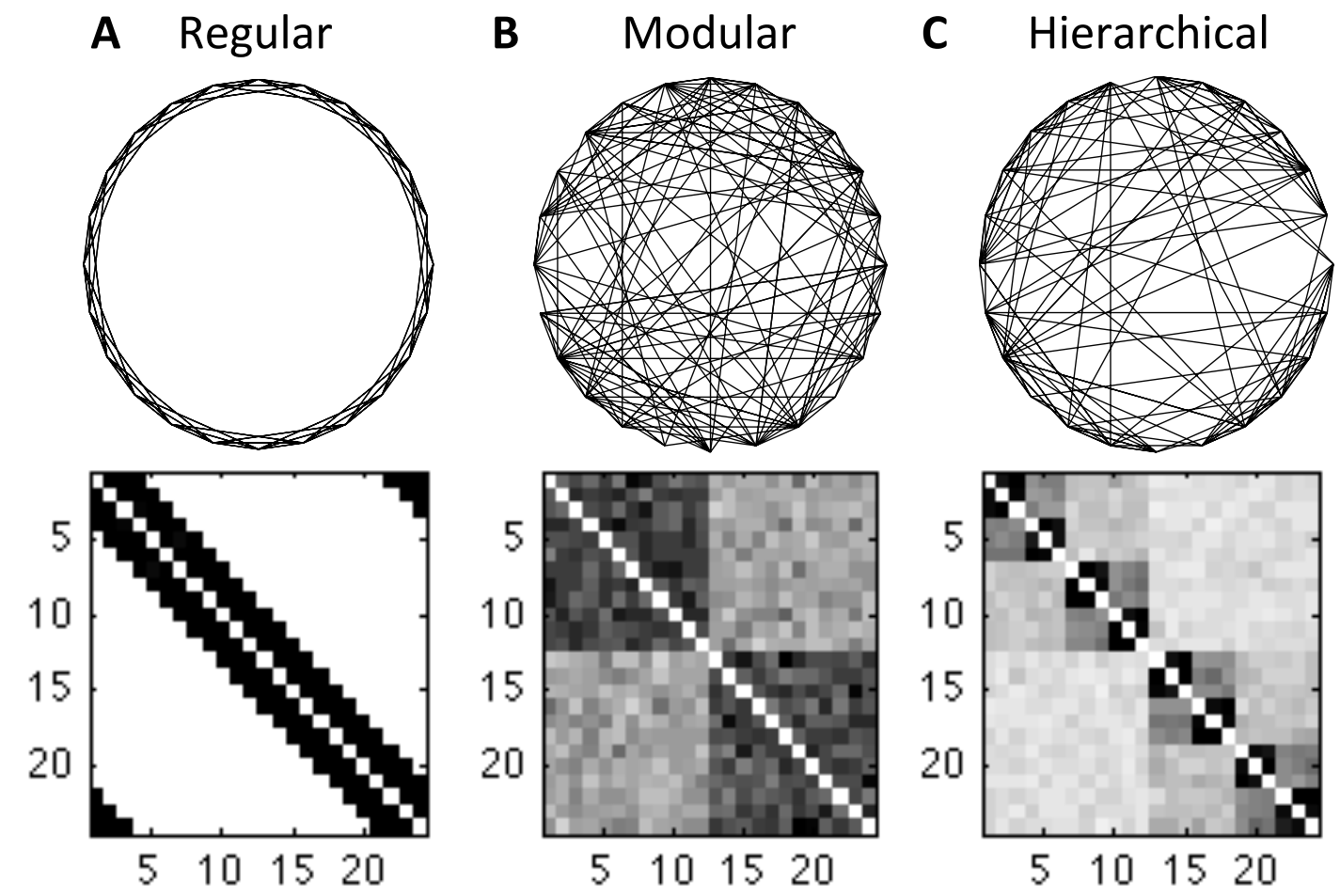

**Fig. 2.** Examples for different types of neural networks. (A) Regular or lattice network. (B) Modular network with two modules. (C) Hierarchical network with two modules consisting of two sub-modules each. Each network contains 24 nodes and 72 bi-directional connections (top: circular arrangement placing nodes with similar neighbours close to each other, thus visualizing modules if present in the network; bottom: average connection frequency for 100 networks of respective type with colour range from black for edges that occur all the time to white for edges that never occur).

**The *Drosophila* connectome**

Two recent studies provide first insights into the connectome of the fruit fly *Drosophila melanogaster* [15,16]. This organism's central brain has around 30,000 neurons and 41 compartments. These spatial compartments are also topological modules. However, nodes that are spatial neighbours are not always in the same topological module and vice versa. For the visual cortex in primates, for example, the frontal eye field is most closely linked to the topological module related to vision, despite many connections to the motor cortex, but not spatially adjacent to other nodes of the visual module. Typically, there is often a link between topological and spatial modules: for the cat cortical network, brain areas within the same topological module also process the same modality [17,18] (Figure 3). There are numerous algorithms to identify modules but their general idea is to find a solution of assigning a module to each node so that the modularity, measured through the modularity value Q [19], is maximized. Solutions show a high number of connections within modules and a low number of connections between modules (see Supp. Note S2 for more information).

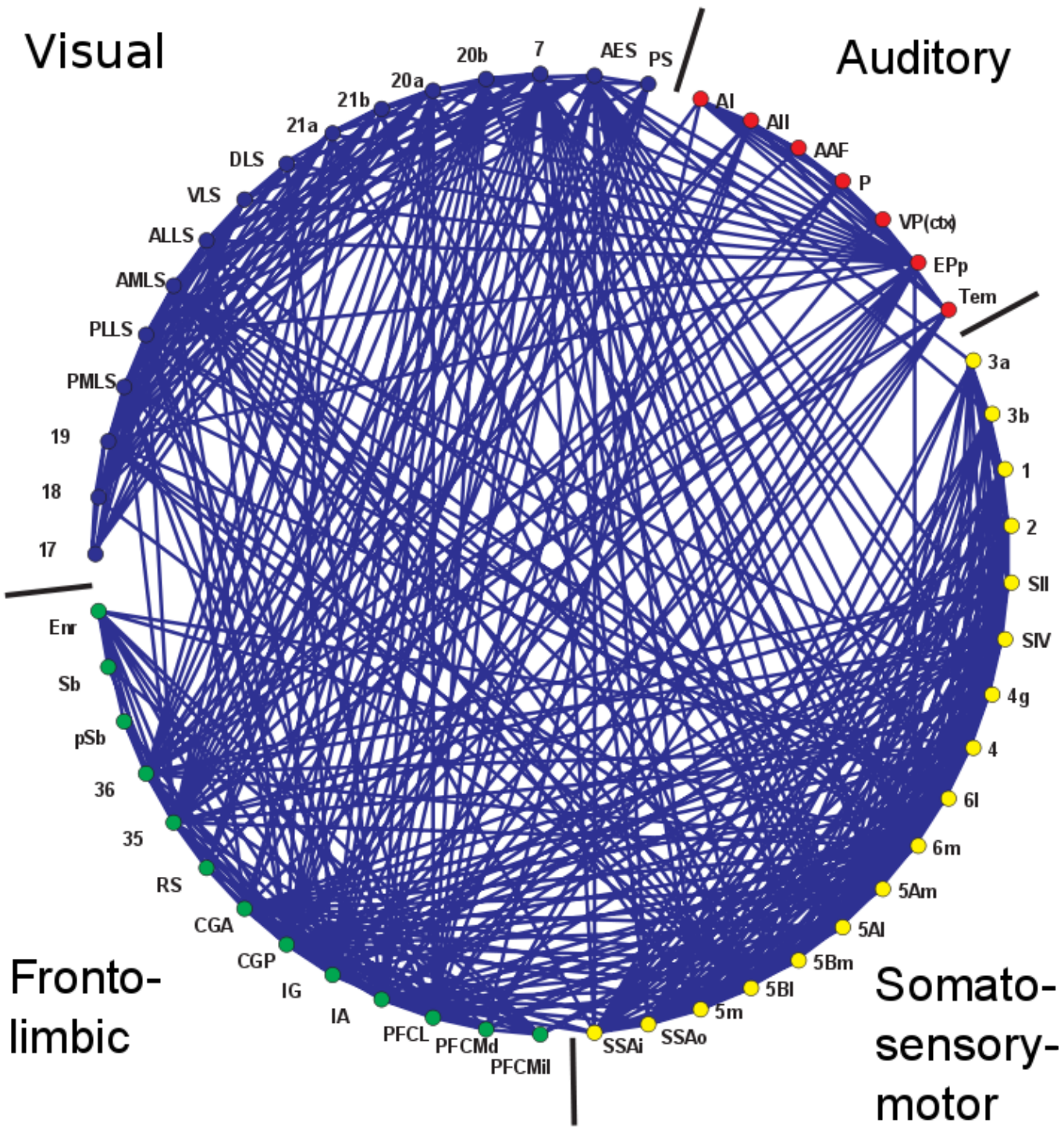

**Fig. 3.** Cluster structure of cat corticocortical connectivity. Bars indicate borders between nodes in separate clusters. Note that nodes in the same cluster, having a high structural similarity, also have a similar function as indicated by node-colours (visual: blue, auditory: red, somatosensory-motor: yellow, and frontolimbic: green).

*Drosophila* also shows a small-world network organisation [16]. A small-world network shows a high connectivity between neighbours such as in a lattice network; the average probability that neighbours (i.e. directly connected nodes) of a node are connected is called the clustering coefficient. In addition, however, small-world networks show 'short-cuts' that link different parts of the network. Using these long-range connections one can quickly reach different parts of the network over few intermediate links; the average number of connections to cross to go from any one node to any other node is called characteristic path length. For small-world networks, the clustering coefficient is much higher than that of randomly wired networks, whereas the characteristic path length is comparable to randomly wired networks (see Supp. Note S2). The values for *Drosophila* are comparable both to neuronal networks in *C. elegans* [20] but also to fibre tract networks in the cat [17] and macaque [18]; all show a small-world organisation despite different levels of brain size and organization. We know that high neighbourhood connectivity results from functional modules but what is the need for short-cuts in all these networks?

**The long way to short-cuts**

Long-distance connections that form short-cuts in a network are expensive in terms of establishment (e.g. myelination) and signal transmission. Neural systems in *C. elegans* and macaque therefore tend to reduce the amount of long-distance connectivity [21,22]. Nonetheless, there are many more long-distance connections in both organisms than would be beneficial; re-arranging the position of nodes could lead to a reduction in total wiring length by 50% and 30%, respectively [23]. Though costly, long-distance connections can reduce the average number of connections in pathways (characteristic path length) leading to faster information processing, higher reliability, and facilitated synchronization [23]. Thus, a lack of long-distance connectivity may lead to cognitive deficits. Indeed, a reduced amount of long-distance connectivity was found for disorders that lead to cognitive deficits such as schizophrenia and epilepsy as well as for Alzheimer's disease.

Long-distance connections are more difficult to establish than short-distance ones. In *Drosophila*, the probability of synaptic contact between two neurons increases with the spatial overlap between the neurites of both neurons [15]. Similarly, the frequency with which two cortical neurons are connected decays exponentially with the distance between them (see [24] for a description of the mechanism and further examples). Establishing a long-distance connection to specific targets needs guidance cues for the axonal growth cone. However, a recent study in *C. elegans* indicates that up to 70% of long-distance connections could be formed early during development when *C. elegans* has only 20% of its adult length [25]. In this way, guiding an axon over a long-distance is not necessary.

**Network robustness towards lesions**

Modules and long-distance connections are two examples for the non-uniform organisation of neural systems. Neural network are also heterogeneous in that some neurons have significantly more connections than others. These nodes with a large number of connections—a large degree—are called hubs. For *Drosophila*, the centro-posterior intermediate (CPI) compartment is a hub [16] whereas sub-cortical regions such as hippocampus and amygdala are the most highly-connected nodes of the macaque network [26]. Interestingly, in all systems, hubs are in the centre of the brain, forming early during development, and presumably originating earlier during evolution. There is mounting evidence that the time that is available for connection establishment, from node formation to brain maturation, plays a crucial role in the establishment of hubs both in vertebrates as well as in *C. elegans* [25].

The synaptic and axonal structure of neural networks is constantly changing in order to adapt to both an external, constantly changing environment as well as to internal changes through learning. Robustness against lesions can therefore be seen as a side effect of adapting to these changes. How does a neural network react to the removal of nodes or edges? Networks that contain hubs are vulnerable towards removal or knockout of these hubs. On the other hand, random removal of nodes will, on average, pick nodes with few connections potentially leading to a smaller deficit after removal [26]. At the same time, both in *Drosophila* and in other neural networks hubs tend to be connected to each other (high assortativity coefficient) leading to an increased resilience towards lesions [16]. Hubs can be crucial for integrating information when they have a high in-degree (number of afferent connections) but can also be important for distributing information when they have a high out-degree (number of efferent connections). For cat and macaque, hubs form a spatially distributed module with the capacity to integrate multisensory information [27].

## Conclusion

Neural systems show a modular and typically also a hierarchical organisation across different levels and across different species (it remains to be seen whether *Drosophila* already shows a hierarchical or only a modular organisation). We mentioned that topology relates to function, but it is also influences dynamics as earlier studies showed its effect on synchrony, oscillation, and activity propagation. Understanding the link between the hierarchical organisation and processing (e.g. does consciousness structurally correlate with the top level of the hierarchy and where is the 'top' in a network?) remains one of the main challenges of the field. In addition, although neuron nodes are often treated as uniform entities, they can differ in terms of function (e.g. inhibitory vs. excitatory), morphology, or gene expression pattern. Whereas heterogeneity concerning links (e.g. hub nodes vs. regular nodes) is well-studied, these node differences deserve more investigation. Finally, this Primer provides only a first glimpse of the emerging field of connectome analysis. Please refer to additional tutorials [28], overviews [29], and toolboxes [30] for more information.

## Acknowledgements

M.K. was supported by WCU program through the National Research Foundation of Korea funded by the Ministry of Education, Science and Technology (R32-10142), the CARMEN e-Science project (www.carmen.org.uk) funded by EPSRC (EP/E002331/1), and (EP/G03950X/1).